\documentclass[sigconf]{acmart}
\usepackage{algorithmicx}
\usepackage[noend]{algpseudocode}  
\usepackage{booktabs}
\usepackage{multirow}
\usepackage[ruled,vlined,linesnumbered,boxed,commentsnumbered]{algorithm2e}

\AtBeginDocument{%
  }

\acmConference[]{}{}{}
  
\begin{document}

\title{Privacy-Preserving Encrypted Low-Dose CT Denoising}

% \author{Anonymous}
% \affiliation{
% \institution{Anonymous Address
% \country{Anonymous Country}}}


\author{Ziyuan Yang}
\email{cziyuanyang@gmail.com}
\affiliation{
\institution{College of Computer Science, Sichuan University}
\city{Chengdu}
\country{China}
}

\author{Huijie Huangfu}
\email{huangfu@stu.scu.edu.com}
\affiliation{
\institution{College of Computer Science, Sichuan University}
\city{Chengdu}
\country{China}
}

\author{Maosong Ran}
\email{maosongran@gmail.com}
\affiliation{
\institution{College of Computer Science, Sichuan University}
\city{Chengdu}
\country{China}
}

\author{Zhiwen Wang}
\email{paprikatree@foxmail.com}
\affiliation{
\institution{College of Computer Science, Sichuan University}
\city{Chengdu}
\country{China}
}

\author{Hui Yu}
\email{smileeudora@163.com}
\affiliation{
\institution{College of Computer Science, Sichuan University}
\city{Chengdu}
\country{China}
}

\author{Yi Zhang}
\authornote{Corresponding Author}
\email{yzhang@scu.edu.cn}
\affiliation{
\institution{School of Cyber Science and Engineering, Sichuan University}
\city{Chengdu}
\country{China}
}
\begin{abstract}
 Deep learning (DL) has made significant advancements in tomographic imaging, particularly in low-dose computed tomography (LDCT) denoising. A recent trend involves servers training powerful models with large amounts of self-collected private data and providing application programming interfaces (APIs) for users, such as Chat-GPT. To avoid model leakage, users are required to upload their data to the server model, but this way raises public concerns about the potential risk of privacy disclosure, especially for medical data. Hence, to alleviate related concerns, in this paper, we propose to directly denoise LDCT in the encrypted domain to achieve privacy-preserving cloud services without exposing private data to the server. To this end, we employ homomorphic encryption to encrypt private LDCT data, which is then transferred to the server model trained with plaintext LDCT for further denoising. However, since traditional operations, such as convolution and linear transformation, in DL methods cannot be directly used in the encrypted domain, we transform the fundamental mathematic operations in the plaintext domain into the operations in the encrypted domain. In addition, we present two interactive frameworks for linear and nonlinear models in this paper, both of which can achieve lossless operating. In this way, the proposed methods can achieve two merits, the data privacy is well protected and the server model is free from the risk of model leakage. Moreover, we provide theoretical proof to validate the lossless property of our framework. Finally, experiments were conducted to demonstrate that the transferred contents are well protected and cannot be reconstructed. The code will be released once the paper is accepted.
\end{abstract}



%
% Keywords. The author(s) should pick words that accurately describe
% the work being presented. Separate the keywords with commas.
% A "teaser" image appears between the author and affiliation
% information and the body of the document, and typically spans the
% page.


%
% This command processes the author and affiliation and title
% information and builds the first part of the formatted document.
\maketitle

%需要额外的一个图
\section{Introduction}
X-ray computed tomography (CT) has gained popularity in clinical diagnosis in recent decades due to its non-invasiveness, high efficiency, and high resolution~\cite{shan2019competitive}. However, concerns regarding the potential health risks associated with the ionizing radiation involved in the scanning process have arisen. To reduce radiation exposure to patients, low-dose technologies has garnered significant interest in recent years, such as switching the voltage/current of X-ray tube and reducing the scanning views~\cite{xia2021ct}. Unfortunately, these methods inevitably result in a degradation in image quality, which can have a substantial negative impact on the subsequent diagnosis~\cite{xia2022low}.

In the past few decades, researchers have made great efforts to develop compressive sensing (CS)-based methods for reconstructing normal-dose computed tomography (NDCT) from low-dose measurements. Generally, these methods usually rely on handcrafted regularization terms derived from prior knowledge and assumptions. However, they usually suffer from high computational costs and laborious parameter adjustments, which limit their applications in clinical practice.

Recently, deep learning (DL) has demonstrated remarkable success in various computer vision tasks and has garnered significant attention for its potential applications in low-dose computed tomography (LDCT) reconstruction. Numerous approaches have been proposed in this field~\cite{xia2023physics}. However, it is widely recognized that DL-based methods rely heavily on large amounts of training data, rendering them data-hungry. In the context of LDCT reconstruction, this presents a challenge as hospitals must adhere to strict ethical and legal obligations regarding patient data confidentiality. As a result, they often face challenges in collecting sufficient data to train powerful DL models, particularly for small hospitals and clinics with limited access to extensive datasets.

An effective solution could be for entities with substantial computational resources, such as major hospitals or CT scanner manufacturers, to collect extensive datasets and train powerful models using their high computational capabilities. Subsequently, these entities could release application programming interfaces (APIs) for user access, similar to the approach demonstrated by Chat-GPT~\cite{openai2023gpt}, which is a notable successful commercial case. This solution comes with multiple advantages as well as certain limitations. On the positive side, it ensures security by effectively preventing any potential leakage of proprietary training data and the model, thereby protecting intellectual property. 
% Furthermore, its user-friendly nature stems from its straightforward and convenient operation. 
Nevertheless, the application of this approach to medical-related tasks faces challenges due to the presence of extensive private patient information in medical data. Hackers may exploit vulnerabilities in the transmission process to gain patients' private data. Thus, it is crucial to address the emerging issue of simultaneously protecting server interests and preserving patients' privacy.

\begin{figure}[t]
    \centering
    \centerline{\includegraphics[width=\columnwidth]{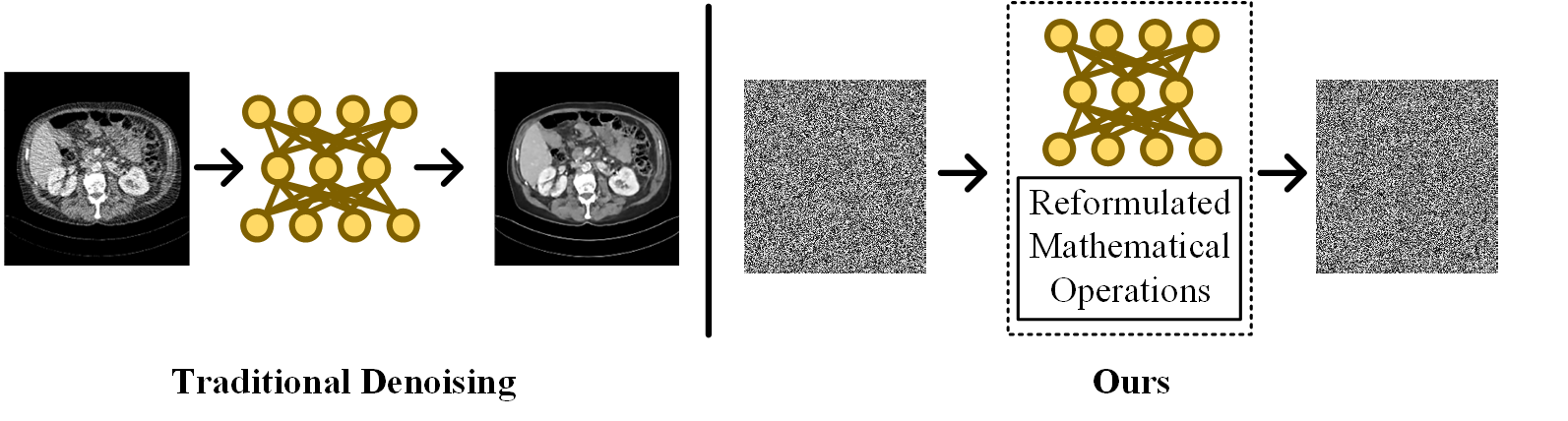}}
    \vspace{-10pt}
    \caption{The concepts of the traditional denoising and ours.}
    \label{fig:concept}
    \vspace{-25pt}
\end{figure}

In this paper, we attempt to address this challenge by developing a novel interactive LDCT image denoising framework that enables users to harness the capabilities of the server model without compromising patient privacy or exposing the server's model. In Fig.~\cite{fig:concept}, two toy examples are provided to illustrate the distinction between traditional denoising interactive frameworks and ours. In the traditional framework, plaintext LDCT images were uploaded to the server for subsequent denoising, which neglects data privacy protection. In contrast, our emphasis is on protecting data privacy without any requirement for retraining the model. We only need to reformulate the mathematical operations of the model to achieve direct denosising the encrypted LDCT images in the encrypted domain.
Specifically, we don't impose any restrictions on the server's model training phase. The server can follow the regular training process to train a model with self-collected plaintext CT images and then deploy it within its infrastructure. Subsequently, users encrypt their data in local and transmit the encrypted LDCT images to the server. After the server processes the encrypted images, the data are returned to the users and then the users decrypt the data to obtain the denoised images.

The procedure is simple and efficient, but the implementation is difficult due to the encrypted data processing in the network model. The conventional convolutional operations and the linear transformation operations, two main components of DL, cannot directly work in the encrypted domain. To circumvent this obstacle, we reformulate the traditional mathematical operations based on homomorphic encryption technology and propose a novel interactive framework to protect the patient's privacy information and the server model. Meanwhile, we mathematically prove the lossless property of our method to support that our reformulated operations in the ciphertext can get equal results with related conventional operations in plaintext.

Activation function is one of the reasons that DL methods can achieve remarkable performance. The activation functions are basically non-linear, and to our best knowledge, it is impossible to find a function to map them to the encrypted domain. As a result, we propose two distributed interactive frameworks for linear and nonlinear DL models, respectively. In the case of linear models, the entire inference process performs exclusively within the server. During a single usage instance, the user's interaction with the server is limited to two times: uploading encrypted LDCT images and subsequently downloading encrypted denoised images.

For nonlinear models, we devise a method to activate features based on the ciphertext without compromising plaintext security. The basic operation of activation functions generally relies on the comparison between the features and a predefined value, such as 0 in the family of ReLu. To this end, we propose an innovative approach to deal with this. The server initiates the process by randomly generating perturbance matrices to the immediate features at first. These matrices are then used to perform dot products with the corresponding encrypted features. Subsequently, users download and decrypt the perturbed features from the server, following which they extract and transmit the binary sign matrix of the outcomes back to the server. Finally, the server can compute lossless activation results by factoring in both the sign matrix of the perturbance matrix and the user-uploaded sign matrix. 

The main contributions of this paper can be summarized as follows:
% \vspace{-\topsep}
\begin{itemize}%[topsep=0px,partopsep=0px]
    \item In this paper, we propose to directly denoise LDCT in the encrypted domain. To the best of our knowledge, this is the first work in this field and we give a novel solution for privacy-preserving LDCT denoising.
    \item We reformulate the key operations in DL to ensure they can work in the encrypted domain and achieve lossless results. We further prove the correctness of the reformulation regarding losslessness.
    \item We present two novel interactive frameworks for linear and nonlinear models, respectively. Additionally, we conduct experiments to verify the effectiveness of our approach in defending against reconstruction attacks.
\end{itemize}
% \vspace{-\topsep}
% The rest of this paper is organized as follows. We briefly review related works in Sec.~\ref{sec:re_wo}, and detailed methods are introduced in Sec.~\ref{sec:meth}. Experiments are shown and discussed in Sec.~\ref{sec:exp}. Finally, we conclude this work and give future works in Sec.~\ref{sec:con}.

% \vspace{-15pt}
\section{Related Works}
\label{sec:re_wo}

\subsection{LDCT Reconstrucction Methods}
With increasing concerns about radiation dose, researchers are dedicated to exploring innovative approaches that can simultaneously reduce radiation dose while maintaining image quality. Many works have been proposed recently to achieve LDCT reconstruction, which can be roughly divided into three categories, namely sinogram filtration, iterative reconstruction (IR), and post-processing.

Sinogram filtering methods directly work on the raw data, and then reconstruct images with analytical methods, such as filtered backprojection (FBP)~\cite{pan2009commercial}. However, raw data is usually difficult to get access from the commercial CT scanners, which limits the promotion of this method. For most IR methods, the reconstruction problem is typically formulated as an optimization problem combining with the prior knowledge in sinogram and/or image domains, such as total variation (TV)~\cite{tv}, non-local means filter~\cite{chen2009bayesian} and some other regularization terms~\cite{chen2013improving, tightFrame, dictlearn}. While these methods achieve impressive performance, the design of regularization terms is often handcrafted, resulting in poor robustness and limited generalization. Besides, these methods are usually computationally expensive and need to adjust parameters to the target image.

Encouraged by the great success of DL, fruitful works about LDCT reconstruction have been proposed in recent years~\cite{li2022noisema, Niu2022Noise, wu2021deep}. Typically, Chen \textit{et al.}~\cite{chen2017low} proposed to introduce the residual connection into autoencoder for LDCT reconstruction, dubbed RED-CNN, which can effectively suppress the noise and reconstruct high-quality CT images. Yang \textit{et al.}~\cite{yang2018low} combined the generative adversarial network (GAN) with Wasserstein distance and perceptual loss. Zhang \textit{et al.}~\cite{zhang2021clear} used multi-level consistency loss to optimize GAN, and named this method as CLEAR. Moreover, some researchers introduce the measured data into the reconstruction process and proposed iteration unrolling-based methods~\cite{xia2023physics}. For example, Chen \textit{et al.}~\cite{chen2018learn} used a convolutional neural network (CNN) to learn an optimal regularization term from training data. Inspired by this work, Xia \textit{et al.}~\cite{xia2022transformer} proposed to combine Transformer and CNN as the regularization term to achieve promising results. 
Xiang \textit{et al.}~\cite{xiang2021fista} combined the model-based Fast Iterative Shrinkage/Thresholding Algorithm (FISTA) into a deep network, dubbed as FISTA-Net. Wang \textit{et al.}~\cite{wang2022uformer} combined Transformer into the U-shaped network and proposed UFormer to denoise LDCT images. Guan \textit{et al.}~\cite{guan2022generative} proposed an unsupervised score-based generative model in the sinogram domain for sparse-view CT reconstruction.
However, the aforementioned methods often require hospitals to collect and upload a substantial amount of private data to the server for training powerful models, without adequately addressing the issue of privacy. This limitation impedes their application and deployment in clinical practice.

\subsection{Privacy-Preserving Methods}

To address the risk of sensitive medical information leakage, some researchers attempt to distributedly train the LDCT models without sharing data. For example, Yang \textit{et al.}~\cite{yang2022hypernetwork} designed physical-driven physical hypernetworks to modulate the global-shared imaging network to avoid data sharing between different parties. Moreover, DC-SFL~\cite{yang2023dynamic} combined the merits of split learning and federated learning to reconstruct LDCT images without sharing local models and local data. Li \textit{et al.}~\cite{li2023semi} proposed a semi-centralized network for LDCT imaging. However, many works have proved that the private data can be reconstructed from the transferred gradients~\cite{huang2021evaluating}. One possible way to alleviate this problem is to introduce encryption technologies in the transmission process to avoid transferring plaintext between parties. For example, Pedrouzo-Ull \textit{et al.}~\cite{pedrouzo2016image} attempted to denoise the encrypted image by using direct and inverse wavelet transform, but this method cannot support DL-based denoising methods. CryptoNet~\cite{gilad2016cryptonets} is an early attempt, but it modifies the activation function, which means it requires servers to retrain their models following their rules, activating features by squaring them. However, this uncommon activation function is unable to distinguish positive and negative features, potentially resulting in gradient vanishing and explosion problems.
Although directly denoise the LDCT images in the encrypted domain appears sufficient safety, to our best knowledge, there is no public work in this field. In this paper, we propose two distributed frameworks to achieve plug-and-play denoise in the encrypted domain for linear and nonlinear DL methods without exposing server models and leaking private data.

\begin{figure*}
% \vspace{8pt}
         \begin{minipage}[t]{0.24\textwidth}
	\centering
	\includegraphics[width=\textwidth]{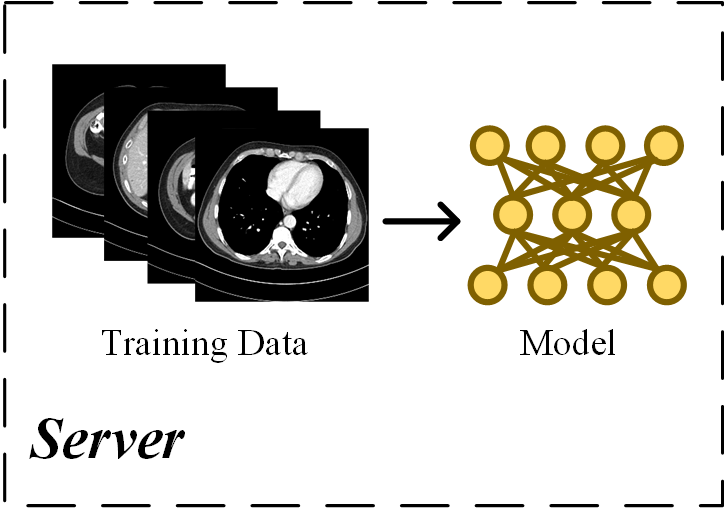}
        % \vspace{15pt}
	\centerline{(a) Training Phase}
	\end{minipage}
          \begin{minipage}[t]{0.75\textwidth}
	\centering
	\includegraphics[width=\textwidth]{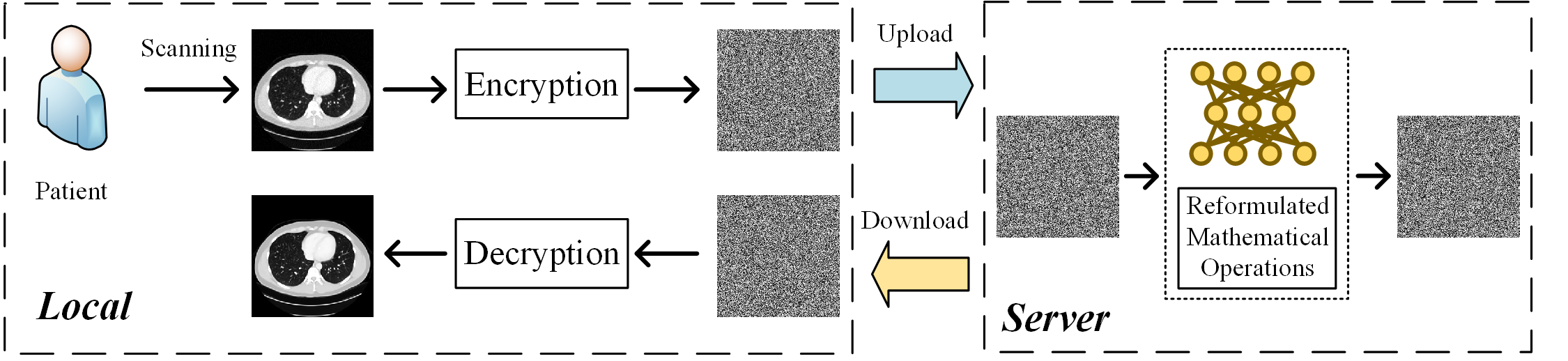}
        % \vspace{15pt}	
        \centerline{(b) Inference Phase}
	\end{minipage}
    % \centering
    % \centerline{\includegraphics[width=\textwidth]{./imgs/framework.png}}
    % \caption{The concepts of the proposed framework. (a) indicates the training phase. The server should train a powerful model with self-collected data. (b) indicates the inference phase. Users are scanned with some low-dose technologies, and then encrypt and upload the the scanned LDCT images to the server. In the server, the model would directly work in the ciphertext based on our reformulated mathematical operations. Finally, the users would decrypt the downloaded encrypted denoised images from the server.}
    \caption{The concepts of the proposed framework. (a) indicates the training phase. (b) indicates the inference phase, patients are firstly scanned with low-dose technologies, and then encrypt and upload the scanned LDCT images to the server. In the server, the model would directly work in the ciphertext based on our reformulated mathematical operations. Finally, the users would decrypt the downloaded encrypted denoised images from the server.}
    \vspace{-5pt}
    \label{fig:concept_framework}
\end{figure*}

\section{Methodology}
\label{sec:meth}
% In this section, we will begin with a problem statement followed by an overview of the proposed framework. Next, we will delve into the reformulation of mathematical operations in the encrypted domain. We will then introduce two interactive framework implementations targeting linear and nonlinear models, respectively. Lastly, we will provide proof demonstrating the equivalence of our reformulated mathematical operations in the encrypted domain with their related operations in the plaintext domain.

\subsection{Problem Statement}
%CT reconstruction problem can be formulated as follows:
%\begin{equation}
%\underset{x}{\min} \frac{1}{2}\|A x-y\|_{2}^{2}+R(x),\label{eq}
%\end{equation}
%where $\|\cdot\|_{2}^{2}$ and $R(\cdot)$ denote the $L_{2}$ norm and the regularization term, respectively. $R(\cdot)$ is generally designed based on some prior knowledge. $x$ and $y$ represent the reconstructed image and measured data, respectively. $A$ represents the system matrix.

The DL-based LDCT image denoising process can be formulated as:
\begin{equation}\underset{\theta}{\min}\|\mathcal{F}(x_l,\theta)-x_n\|_{2}^{2},\end{equation}
where $x_l$ and $x_n$ denote the LDCT image and corresponding NDCT image, respectively. $\mathcal{F}$ is the objective model parameterized by $\theta$.

In this paper, we attempt to directly denoise the encrypted LDCT images, then the concept of this process is reformulated as follows:
\begin{equation}
    Enc(\hat{x}) = \mathcal{F}(Enc(x_l),\theta),
\end{equation}
where $Enc(\cdot)$ denotes the encryption function, and $\hat{x}$ represents the denoised LDCT image.

The decrypted denoised LDCT image should be equal to $\hat{x}$, which is the sufficient and necessary condition for achieving lossless denoising. The condition can be formulated as:
% it is necessary to ensure that they satisfy the following condition:

\begin{equation}
    Dec(Enc(\hat{x})) = \hat{x},
\end{equation}
where $Dec(\cdot)$ represents the decryption function.

% Subsequently, we will discuss the main steps involved in the encryption and decryption processes within the encryption cryptosystem. 

\subsection{Overview}
\label{sec:overview}
As mentioned above, existing DL-based LDCT denoising methods have overlooked privacy concerns. Although distributed training-based methods, e.g. Federated Learning (FL) methods, can mitigate this problem to a certain degree, they still face challenges, due to the frequent transmission of gradients during the training phase between different parties, which poses a dual risk of reconstruction attacks and model parameter leakage. To alleviate these concerns, in this paper, we propose a novel framework for privacy-preserving distributed LDCT denoising for DL-based methods.

The concept of the proposed framework is illustrated in Fig.~\ref{fig:concept_framework}. For the training phase, there is no requirement to modify the training setting. The server can adhere to its original settings and train its powerful model with a large amount of self-collected data. Our proposed framework is plug-and-play, which can be directly embedded into trained models without any extra operations. For the inference phase, the users encrypt their LDCT images locally and transfer them to the server. The server uses the trained model equipped with our reformulated mathematical operations to denoise the encrypted LDCT images in the encrypted domain. Finally, the users download the denoised results from the server and decrypt them to obtain the predicted NDCT images. Crucially, the server side remains devoid of any decryption procedure, guaranteeing that the plaintext remains inaccessible to the server. Simultaneously, users are unable to access the server model, significantly reducing the potential risk of model parameter leakage. Consequently, it creates a win-win situation for both users and the server. The server can confidently release its APIs to users without worrying about model leakage, while users can be assured that their data will not be reconstructed or compromised.

\subsection{Homomorphic Encryption}
Homomorphic encryption technologies are widely used to ensure security in various applications. The key advantage of a homomorphic cryptosystem lies in its ability to achieve lossless mathematical operations in the encrypted domain. By leveraging homomorphic encryption, it is possible to perform arithmetic operations directly on encrypted data. In this paper, we utilize the Paillier encryption scheme~\cite{paillier1999public} to achieve privacy-preserving, which is also widely used in distributed learning methods. Here we briefly introduce the encryption and decryption processes in our framework based on Paillier.
% To provide an intuitive overview of the overall process, we will begin by introducing the encryption and decryption operations.

As introduced in Section~\ref{sec:overview}, the users should encrypt their data at first. Specifically, users should first generate public and private keys locally as follows:
\begin{equation}
\label{eq:key}
    n = p \cdot q,
\end{equation}
where $p$ and $q$ are two different large prime numbers that satisfy $\mathrm{gcd}(pq,(p-1)(q-1))=1$, where $\mathrm{gcd}(a,b)$ denotes the greatest common divisor of $a$ and $b$. Subsequently, we can arbitrarily choose a random number $g \in \mathbb{Z}_{n^2}^*$ such that $\mathrm{gcd}(L(g^\lambda \bmod n^2), n) = 1$, where $L(u)= \frac{u-1}{n}$, and $\lambda= \mathrm{lcm}(p-1,q-1)$ is the private key. $\mathrm{lcm}(a,b)$ denotes the least common multiple of $a$ and $b$. The public key is $(n,g)$

Once the public and private keys are generated, the users encrypt their LDCT images before transmitting them to the server. Specifically, users should randomly select a random number $r$, which satisfies that $0<r<n$ and $\mathrm{gcd}(r,n)=1$. Then, the ciphertext can be obtained as follows:
\begin{equation}
\label{eq:encry} 
c = g^m \cdot r^n \bmod n^2,
\end{equation}
where $c\in\mathbb{Z}^*_{n^2}$ is the ciphertext, and $m$ denotes unencrypted value, which can be considered as the pixels' values in the plaintext LDCT images or the plaintext features. For convenience, we introduce $Enc(\cdot)$ to represent the encryption process described in Eq.~\eqref{eq:encry}.

After encryption, the users should upload the ciphertext to the server for further processing. In this way, even if the hacker intercepts the uploaded data, they still cannot decrypt it without the private key, which is preserved locally. 
% Because it is impossible to decrypt the ciphertext without the private key, which is preserved in local. 

Upon receipt of the ciphertext, the server treats the ciphertext as input of the trained model. However, it is evident that conventional mathematical operations cannot be directly applied in the encrypted domain. As a result, a reformulation of these mathematical operations is essential to ensure lossless calculations.
To maintain clarity in explaining the entire encryption and decryption processes, the details of this reformulation are given in Sec.~\ref{sec:refor}. Within the confines of this subsection, we proceed with the assumption that we have successfully obtained the encrypted denoised LDCT image, denoted as $Enc(\hat{x})$, and subsequently transferred it to the user.
Once the user receives the encrypted denoised CT image, the decryption stage begins as follows:

% There are some important details about the mathematical operations in the encrypted domain, but for ease of understanding the whole framework, we assume that we've gotten $Enc(\hat{x})$, the encrypted denoised LDCT image, and then transfer it to the user. Finally, users would decrypt the ciphertext to get the denoised plaintext CT images as:
\begin{equation}
\label{eq:dec}
    \hat{x} = L(c^\lambda \bmod n^2) \cdot \mu \bmod n,
\end{equation}
where $\mu$ is the modular inverse of $L(g^\lambda \bmod n^2)$ modulo $n$, $\hat{x}$ refers to the decrypted value. We then introduce the the notation of decryption process as $Dec(\cdot)$ in the following descriptions.

In the whole process, the private key is not transferred and preserved locally. Since only the user can decrypt the encrypted CT images, the privacy information contained in the data is well preserved in the process. Additionally, since the server model is not transferred to other parties, it is also well protected.

\subsection{Reformulation for the Mathematical Operations in the Encrypted Domain}
\label{sec:refor}

As indicated in the previous subsection, conventional mathematical operations cannot be directly executed within the encrypted domain. Consequently, it becomes necessary to develop an alternative that ensures the denoising model can effectively perform in the encrypted domain while maintaining lossless.

Regardless of whether the denoising model is CNN-based ~\cite{chen2017low} or Transformer-based~\cite{wang2022uformer}, the fundamental mathematical operations shared between them predominantly contain addition and multiplication operations. For example, the convolutional operation is formulated as:
\begin{equation}
    f * g(u, v)=\sum_{i} \sum_{j} f(i, j) g(u-i, v-j),
\end{equation}
where $g$ represents the image or feature, and $f$ represents the convolutional filter. $(u,v)$ denotes coordinate index of $g$, and "$*$" denotes the convolutional operation.

To simplify this process, we assume that $f(x, y)=a_{x, y}$ and $g(x, y)=b_{x, y}$, then the above function can be expressed as:
\begin{equation}
\label{eq:conv}
   f * g(u, v)=\sum_{i} \sum_{j} a_{i, j} b_{u-i, v-j}.
\end{equation}

It is easily found that in Eq.~\eqref{eq:conv}, the convolution operation can be viewed as a combination of multiple addition and multiplication operations. Similarly, linear layers and different matrix operations can also be treated as different forms of the same fundamental mathematical operations. Since the goal of this paper is to achieve denoising in the encrypted domain, we need to reformulate the main mathematical operations, such as addition and multiplication, from operating on plaintext to ciphertext.

%%%%20230903

% As mentioned earlier, most DL operations in the encrypted domain only involve addition and multiplication.
The Paillier encryption cryptosystem exhibits homomorphic additive and scalar multiplication properties, which make it possible to reformulate the mathematical operations into the encrypted domain. In this subsection, we only provide the reformulated results and the related proof is given in \ref{sec:Proof}.
% The Paillier encryption cryptosystem, which exhibits homomorphic additive and scalar multiplication properties, is utilized to reformulate the mathmeatical operations into the encrypted domain. In this subsection, we will directly provide the reformulation without proof, and the related correctness proof can be found in  \ref{sec:Proof}.

In this paper, we assume that the parameters of the DL model remain unencrypted, while the input is encrypted to protect privacy. Besides, the model is trained with unencrypted CT images. As mentioned earlier, most mathematical operations, such as Eq.~\eqref{eq:conv}, in DL can be dissected into two fundamental operations: a). the addition of one encrypted value to another encrypted value; and b). the multiplication of an unencrypted value by an encrypted value. In this scenario, the encrypted value could be the feature and the unencrypted value could be the parameters.

Assume that the ciphertext $c_{m_1}$ corresponds to plaintext $m_1$, and similarly, $c_{m_2}$ corresponds to the plaintext $m_2$. Then, performing a homomorphic addition operation on these ciphertexts yields the ciphertext $c_{add}$ that corresponds to the plaintext $m_{add}$, where $m_{add}=m_1+m_2$. Consequently, the addition operation in the encrypted domain can be formulated as follows:

% Then, the homomorphic addition operation on these ciphertext results in ciphertext $c_{add}$ corresponding to plaintext $m_{add}$, where $m_{add}=m_1+m_2$. Then, the addition operation in the encrypted domain can be formulated as:

\begin{equation}
    c_{add} = (c_{m_1}\cdot c_{m_2}) \bmod n^2,
    \label{eq:add}
\end{equation}
where $c_{add}$ represents the addition result in the encrypted domain. Due to the homomorphic additive property in Paillier, the user can decrypt the ciphertext $c_{add}$ to get the protected information. Meanwhile, the decrypted $c_{add}$ is euqal with $m_{add}$, which is defined as $Dec(c_{add})=m_{add}$.

In this way, we can reformulate the addition operation from the plaintext domain into the encrypted domain. As mentioned earlier, the server model's parameters are not encrypted. Hence, another key operation is the multiplication of a plaintext scalar by an encrypted value. Then, the multiplication operation is reformulated as follows:

\begin{equation}
     c_{mul} = c_{m_1}^a \bmod n^2,
     \label{eq:enc_mul}
\end{equation}
where $a$ is the unencrypted scalar, and $c_{mul}$ is the multiplication result in the encrypted domain. In our framework, $a$ can be considered as the model parameter. Similar to the reformulated addition operation, the user can decrypt $c_{mul}$ to obtain $m_1\cdot a$ aided by the homomorphic multiplication property, which is formulated as $Dec(c_{mul})=m_1\cdot a$.

In this way, we can use Eqs. \eqref{eq:add} and \eqref{eq:enc_mul} to reformulate most operations in the plaintext domain. For instance, we could use them to reformulate the convolution operation in~\eqref{eq:conv}, ensuring its feasibility in the encrypted domain. The reformulated operation can be expressed as:
\begin{equation}
    f * Enc(g(u, v))=\prod_{i} \prod_{j}(Enc(b_{u-i, v-j})^{a_{i, j}} \bmod{n^2})\bmod{n^2}.
\end{equation}

By reformulating the mathematical operations for ciphertext, it is possible for the denoising models to efficiently perform LDCT denoising in the encrypted domain, all while bypassing the requirement for decryption on the server side. In light of the fact that the reformulated functions represent fundamental mathematical operations, their combinations make the mapping of various DL operations into the encrypted domain possible. Consequently, our method exhibits a high degree of generality, which can be easily integrated with different DL models. Notably, this integration can be accomplished without any extra operations or modifications to the model architecture.

% By reformulating the mathematical operations for ciphertext, we can effectively perform denoising operations in the encrypted domain without the need for decryption in the server. This approach ensures that the generality of the proposed method, can combined with most DL models without retraining or changing the model architecture.

\vspace{-5pt}
\subsection{Privacy-Preserving Interactive Frameworks}
To provide a clear understanding of the whole process, it is necessary to establish a foundational set of variables. $\mathcal{C}_l$ represents the encrypted LDCT image, and $\mathcal{C}_d$ denotes the denoised encrypted CT image. Besides, we assume that the denoising model $\mathcal{F}$ consists of $k$ layers. Its $i$-th layer is represented as $\mathcal{F}^i$, and its corresponding input is denoted as $I^i$.

Although as shown in Sec. \ref{sec:refor}, some mathematical operations can be reformulated from the plaintext domain to the encrypted domain, some other operations, which cannot be expressed as a combination of addition and multiplication operations, cannot be reformulated in the encrypted domain. The conversion of nonlinear operations, in particular, presents challenges within the context of encryption. Hence, we introduce two distributed solutions tailored to two distinct scenarios: (1) when the model does not encompass any nonlinear operations, and (2) when nonlinear operations are included within the model.

% \begin{algorithm}[t]
% % \SetAlgoLined %显示end
% % \vspace{4pt}
%   \caption{Main steps of the proposed interactive framework for linear models.}  
%   \label{alg:Linear}
% %   \SetKwFunction{CompSer}{CompMain}
% %   \SetKwProg{Fn}{Function}{}{}
%    \textbf{Input:} The scanned LDCT image $x_l$, and the server collected dataset $\mathcal{D}$.\\
%     \textbf{Output:} The denoised LDCT image $\hat{x}$.\\
%    % \vspace{3pt}
%    \textbf{Function Main:} \Comment{Server Executes}\\
%     Train the reconstruction model $\mathcal{F}$ in $\mathcal{D}$. \\ 
%     Release API to users. \\
%     $\mathcal{C}_l \gets$ \textbf{UserEncrypt()} \\
%     $\mathcal{C}_d \gets \mathcal{F}(\mathcal{C}_l)$  \Comment{Eqs. \eqref{eq:conv} and \eqref{eq:add}} \\
%     \textbf{UserDecrypt}($\mathcal{C}_d$) \\
%     % \vspace{3pt}
%     \textbf{Function UserEncrypt():} \Comment{User Executes} \\
%     Generate public key $n$ and private key $\lambda$.%\\
%     \Comment{Eq.~\eqref{eq:key}}\\
%     Generate encrypted LDCT image $\mathcal{C}_l$ from $x_l$.%\\
%     \Comment{Eq.~\eqref{eq:encry}}\\
%     return $\mathcal{C}_l$\\
%      % \vspace{3pt}
%      \textbf{Function UserDecrypt}($\mathcal{C}_d$): \Comment{User Executes} \\
%     Decrypt $\mathcal{C}_l$ and get $\hat{x}$. \Comment{Eq.~\eqref{eq:dec}}%\\
%     % \vspace{4pt}
% \end{algorithm}
% % \vspace{-30pt}

\subsubsection{Interactive Framework for Linear Models}

In this subsection, we assume that the neural network does not contain any nonlinear operations. Under this assumption, all arithmetic operations in the forward propagation can be reformulated as mentioned in Sec.~\ref{sec:refor}. In other words, in the inference phase, all operations are executed on the server side, and no computational requirement is required on the user side. The user is only involved in two data exchange instances with the server in the whole inference process: (1) upload the encrypted LDCT images to the server; and (2) download the encrypted denoised CT images from the server.
% To aid readers in understanding the complete process, the main steps are outlined in Algorithm \ref{alg:Linear}. The data exchange processes can be found in Lines 6 and 9 in Algorithm \ref{alg:Linear}, respectively. Consequently, the overall transfer overhead is considered satisfactory.

\subsubsection{Interactive Framework for Nonlinear Models}

In practice, there are some operations in DL that are nonlinear. One of the most representative nonlinear operations is the activation function. If a neural network lacks nonlinear activation functions, it is unable to effectively learn the nonlinear representation, limited to linear relationships~\cite{apicella2021survey}. However, it is not feasible to reformulate these operations into the encrypted domain.

To achieve nonlinear mapping in the encrypted domain, we introduce a novel interactive framework, which attempts to offload nonlinear operations of the DL model to the user side. There are three key rules: (1) we should protect the privacy of the server model, ensuring that users can compute the activation function without disclosing the features of the model; (2) the server cannot access any plaintext contents in the process; and (3) the activated results should be lossless.

\begin{figure}[t]
    \centering
    \centerline{\includegraphics[width=\columnwidth]{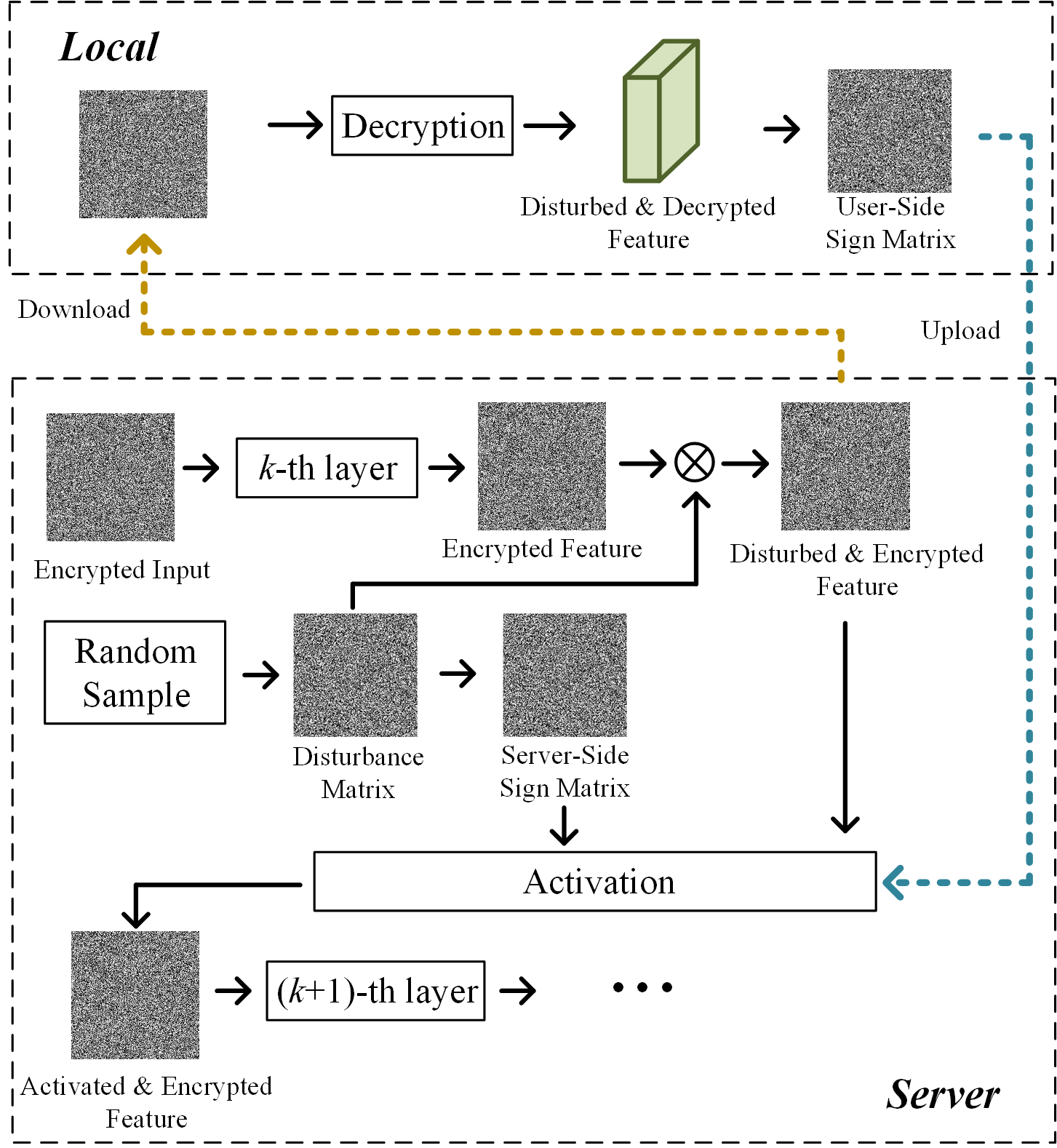}}
    \vspace{-5pt}
    \caption{The visualization of the process of the proposed interactive framework for nonlinear models.}
    \label{fig:inter_act}
\end{figure}

Among different activation functions, the family of ReLU stands out as a popular choice, which has been applied in diverse tasks~\cite{li2017convergence}. Therefore, the primary focus of this paper is the scenarios in which models employ the family of ReLU as the activation functions. We visualize the whole procedure in Fig.~\ref{fig:inter_act}. Basically, different ReLUs utilize 0 as the threshold. For activation, positive and negative features are multiplied by different coefficients. For example, in ReLU, the coefficients are set to 1 and 0 for positive and negative features, respectively. Hence, we just need to know the sign information of features, then the server can activate features in the encrypted domain. The sign function is defined as:
\begin{equation}
    sign(\alpha) = \left\{
    \begin{array}{cl}
    1, & \alpha\ge 0 \\
    0, & \alpha<0 \\
    \end{array}. \right.
\end{equation}

We denote that $\mathcal{C}^i$ is the feature of $i$-th layer, and $Dec(\mathcal{C}^i)=\mathcal{Q}^i$, where $\mathcal{Q}^i$ represents the $i$-th layer decrypted feature. Obviously, it raises concerns about privacy if the server directly transfers $\mathcal{C}^i$ to the user, which means that the users can access all features of different layers and extract the knowledge in the server model. To avoid this, we propose to randomly sample a matrix to perturb features. For $i$-th layer, the randomly sampled perturbance matrix is denoted as $\mathcal{M}^i$, which has the same dimension as $\mathcal{C}^i$. $\mathcal{S}_{s}^i$ stands for the sign matrix of $\mathcal{M}^i$, which is a binary matrix and stored in the server. Based on Eq.~\eqref{eq:enc_mul}, we can obtain the perturbed feature $\mathcal{C}_{per}^i$ as follows:
\begin{equation}
    % \mathcal{C}_{dis}^i = \mathcal{S}_{s}^i \odot \mathcal{C}_i,
    \mathcal{C}_{per}^i(u,v) = \mathcal{C}^i(u,v)^{\mathcal{M}^i(u,v)} \bmod n^2,
    \label{eq:disturb}
\end{equation}
where $(u,v)$ stands for the coordinate index.

Once the user receives $\mathcal{C}_{per}^i(u,v)$ from the server, the user should decrypt it to get $\mathcal{Q}_{per}^i$, where $\mathcal{Q}_{per}^i = \mathcal{M}^i \odot \mathcal{Q}^i$ and $\odot$ denotes the Hadamard Product operation. Due to the lack of $\mathcal{M}^i$, the users cannot access the original features of the model, which can effectively preserve the model privacy. Then, the user should activate the perturbed features as follows:
\begin{equation}
    \mathcal{Q}_{act}^i(u,v) = ReLU(\mathcal{Q}_{per}^i(u,v)),
    \label{eq:relu}
\end{equation}
where $\mathcal{Q}_{act}^i$ denotes the unencrypted and activated feature of ${Q}_{per}^i$. Then, we obtain the sign matrix of $\mathcal{Q}_{act}^i$:
\begin{equation}
    \mathcal{S}_u^i = sign(\mathcal{Q}_{act}^i),
    \label{eq:sign}
\end{equation}
where $\mathcal{S}_u^i$ represents the user-side sign matrix of $\mathcal{Q}_{act}^i$.

%%%% 20230904

\setlength{\textfloatsep}{0.1cm}
\setlength{\floatsep}{0.1cm}
\begin{algorithm}[t]
% \SetAlgoLined %显示end
% \vspace{4pt}
  \caption{Main steps of the proposed interactive framework for nonlinear models.}  
  \label{alg:Nonlinear}
%   \SetKwFunction{CompSer}{CompMain}
%   \SetKwProg{Fn}{Function}{}{}
   \textbf{Input:} The scanned LDCT image $x_l$, and the server collected dataset $\mathcal{D}$.\\
    \textbf{Output:} The denoised LDCT image $\hat{x}$.\\
   % \vspace{3pt}
   \textbf{Function Main:} \Comment{Server Executes}\\
    Train the denoising model $\mathcal{F}$ in $\mathcal{D}$. \\ 
    Release API to users. \\
    $I^1 \gets$ \textbf{UserEncrypt()} \\
    \For{layer $i=1,2,...,k$}{
    $\mathcal{C}^i \gets \mathcal{F}^i(I^i)$  \Comment{Eqs. \ref{eq:conv} and \ref{eq:add}} \\
    Randomly sample a perturbance matrix $\mathcal{M}^i$. \\
    Generate $\mathcal{C}_{per}^i$ and $\mathcal{S}_u^i$ based on $\mathcal{M}^i$.%\\   
    \Comment{Eq.~\eqref{eq:disturb}} \\
    $\mathcal{S}_u^i \gets$\textbf{UserAct}($\mathcal{C}_{per}^i$) \\
    $\mathcal{S}^i = \mathcal{S}_u^i \odot \mathcal{S}_s^i$ \\
    Get the activated encrypted feature $\mathcal{C}_{act}^i$.%\\ 
    \Comment{Eq.~\eqref{eq:disturb}} \\  
    $I_{i+1}\gets \mathcal{C}_{act}^i$
    }
    \textbf{UserDecrypt}($\mathcal{C}_{act}^k$) \\
    % \vspace{3pt}
    \textbf{Function UserEncrypt():} \Comment{User Executes} \\
    Generate public key $n$ and private key $\lambda$.%\\
    \Comment{Eq.~\eqref{eq:key}}\\
    Generate encrypted LDCT image $\mathcal{C}_l$ from $x_l$.%\\
    \Comment{Eq.~\eqref{eq:encry}}\\
    return $\mathcal{C}_l$\\
     % \vspace{3pt}
     \textbf{Function UserAct}($\mathcal{C}_{per}^i$): \Comment{User Executes} \\
    $\mathcal{Q}_{act}^i \gets ReLU(\mathcal{Q}_{per}^i)$ \Comment{Eq.~\eqref{eq:relu}}\\
    $\mathcal{S}_u^i \gets sign(\mathcal{Q}_{act}^i)$ \Comment{Eq.~\eqref{eq:sign}}\\
    return $\mathcal{S}_u^i$ \\
     % \vspace{3pt}
     \textbf{Function UserDecrypt}($\mathcal{C}_d$): \Comment{User Executes} \\
    Decrypt $\mathcal{C}_l$ and get $\hat{x}$. \Comment{Eq.~\eqref{eq:dec}}\\
    % \vspace{4pt}
\end{algorithm}

Then, the user can upload $\mathcal{S}_u^i$ to the server. Since $\mathcal{S}_u^i$ only contains the sign information, it is not feasible to inverse the plaintext features from the ultra-limited prompt. As a result, the plaintext features are also well preserved. Furthermore, $\mathcal{S}_u^i$ is a binary matrix, implying that the associated uploading overhead is acceptable. In this way, the server has the sign information of the perturbance matrix and the perturbed feature, and it is easy to get the sign information of the plaintext features. The process is formulated as:
\begin{equation}
    \mathcal{S}^i = \mathcal{S}_u^i \odot \mathcal{S}_s^i,
\end{equation}
where $\mathcal{S}^i$ denotes the real sign information of $\mathcal{Q}^i$. %Positive and negative values stand for 1 and 0 in $\mathcal{S}^i$, respectively. 
Finally, the server can obtain the activated encrypted feature as:
\begin{equation}
\mathcal{C}_{act}^i(u,v) = \mathcal{C}^i(u,v)^{\mathcal{S}^i(u,v)} \bmod n^2,
\end{equation}
where $\mathcal{C}_{act}^i$ denotes the activated feature, and it is also the input of the next layer, which means $I^{i+1} = \mathcal{C}_{act}^i$.

% 0905   平移即可。

Then we can repeat the above steps until the server obtains the network's final output. Compared to the whole computational cost of the model, the cost of the activation function is almost negligible. Therefore, our proposed framework does not impose extra requirements on users' computing resources. Additionally, the received contents by the users are heavily perturbed and only contain the sign information. It is almost impossible to extract the knowledge of the server model.

Besides, the server only accesses the sign information of the plaintext features and is unable to decrypt the ciphertext. Thus, the proposed framework effectively preserves the private information of both the user and the server model simultaneously. To facilitate the understanding of the proposed interactive framework for nonlinear models, we illustrate the main steps in Algorithm \ref{alg:Nonlinear}. Although the threshold of activation functions is simply treated as 0 in this paper, our proposed framework can be easily extended for other thresholds. Only a simple shift operation is needed to transform the threshold to 0. The user only needs to encrypt and upload the threshold to the server. Then, the server can add this value to the features to keep the threshold to 0.

\subsection{Proof}
\label{sec:Proof}
In this paper, we reformulate two basic mathematical operations in the classic DL models: (1) the addition of one encrypted value to another encrypted value; and (2) the multiplication of an unencrypted value by an encrypted value. Here we validate the correctness of these reformulated mathematical operations. For detailed proofs related to security, encryption, and decryption, we refer readers to~\cite{paillier1999public}. It is important to note that the proof assumes the proper generation and secure management of the cryptosystem parameters, such as $g$, $n$, $r$, and so on.

% https://zhuanlan.zhihu.com/p/420503254

For the first operation, we need to ensure $Dec(c_{add})=m_{add}$. At first, we reformulate Eq.~\eqref{eq:add} as:
\begin{equation}
\begin{aligned}
c_{a d d}&=c_{m_1} \cdot c_{m_2} \bmod n^{2}\\
&=[\left(g^{m_{1}} r^{n}\right)\left(g^{m_{2}} r^{n}\right)] \bmod n^{2} \\
&=g^{m_{1}+m_{2}} r^{n} \cdot r^{n} \bmod n^{2} \\
&=g^{m_{1}+m_{2}} r^{2n} \bmod n^{2},
\end{aligned}
\end{equation}
where $c_{m_1}=Enc(m_{1})=g^{m_{1}}r^{n}\bmod n^{2}$, and $c_{m_2}=Enc(m_{2})=g^{m_{2}}r^{n}\bmod n^{2}$.
% , and $r = r_1 \cdot r_2 \mod n^2$.

Then, we would decrypt $g^{m_{1}+m_{2}} r^{2n} \bmod n^{2}$, which can be expressed as $Dec(g^{m_{1}+m_{2}} r^{2n} \bmod n^{2})$. Since as mentioned earlier, $\mathrm{gcd}(r,n)=1$, we have $(r^2n)\equiv 1 \pmod{n^2}$ based on the Carmichael's theorem~\cite{carmichael1910note}. Hence, the decrypted value is $m_1+m_2$. So far, we have proven the correctness of the reformulated addition operation in the encrypted domain.

% Then, let's decrypt $c_{add}$ as follows:
% \begin{equation}
% \begin{aligned}
% Dec\left(c_{a d d}\right)&=L\left(c_{a d d}^{\lambda} \bmod n^{2}\right) \cdot \mu \bmod n \\
% &=L\left(\left(g^{m_{1}+m_{2}} r^{n}\right)^{\lambda} \bmod n^{2}\right) \cdot \mu \bmod n \\
% &=L\left(g^{\left(m_{1}+m_{2}\right) \lambda} r^{n \lambda} \bmod n^{2}\right) \cdot \mu \bmod n.
% \end{aligned}
% \end{equation}

% Then, the above expression simplifies to:
% \begin{equation}
% \begin{aligned}
%     Dec\left(c_{add}\right)&=L\left(g^{\left(m_{1}+m_{2}\right) \lambda} \bmod N^{2}\right) \cdot \mu \bmod N \\
%     &=\left(m_{1}+m_{2}\right) \bmod N.
% \end{aligned}
% \end{equation}
% Because that $r^{n\lambda}\bmod n^2 = 1$

\begin{figure*}
        \begin{minipage}[t]{0.22\textwidth}
	\centering
	\includegraphics[width=\textwidth]{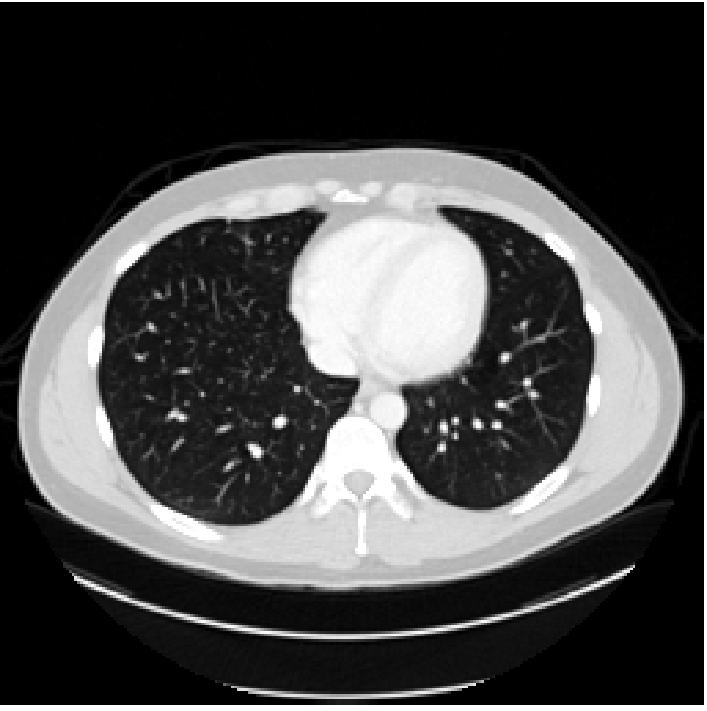}
	\centerline{(a)}
	\end{minipage}
         \begin{minipage}[t]{0.22\textwidth}
	\centering
	\includegraphics[width=\textwidth]{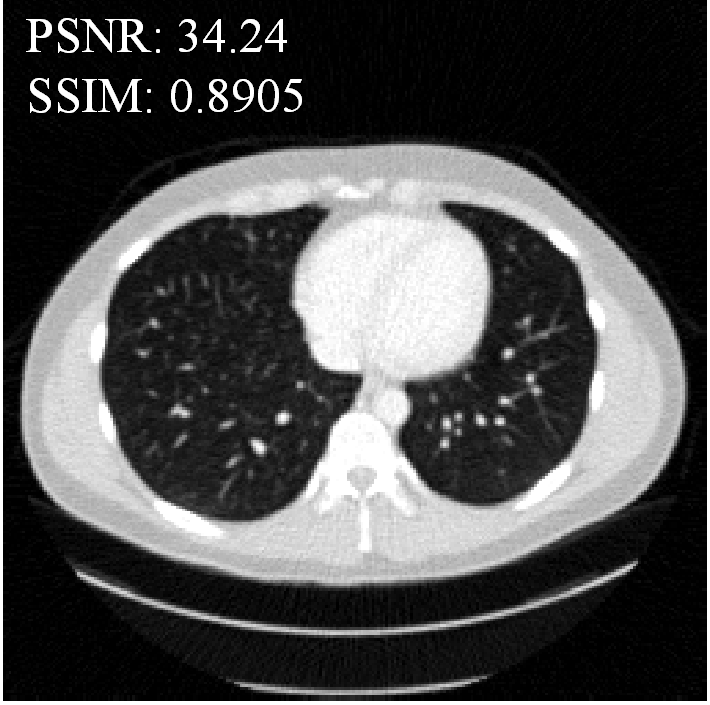}
	\centerline{(b)}
	\end{minipage}
         \begin{minipage}[t]{0.22\textwidth}
	\centering
	\includegraphics[width=\textwidth]{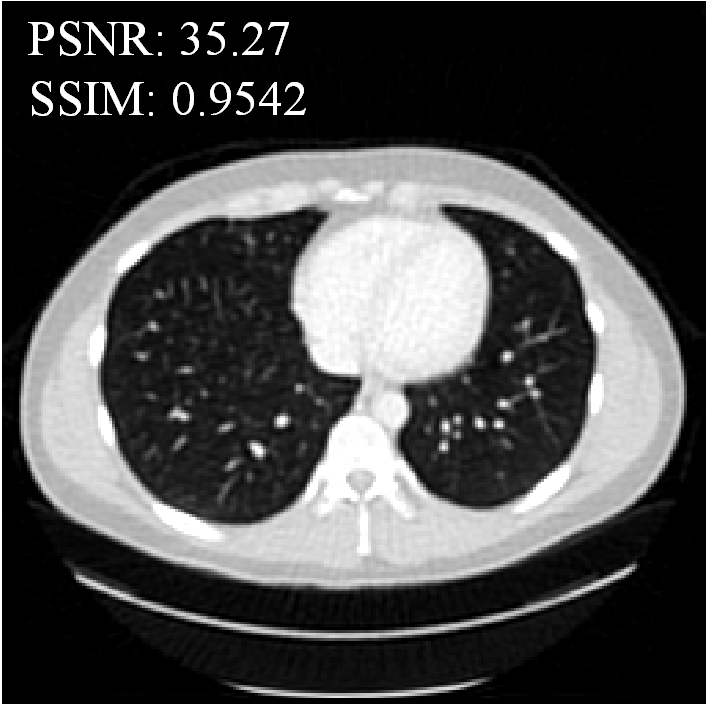}
	\centerline{(c)}
	\end{minipage}
         \begin{minipage}[t]{0.22\textwidth}
	\centering
	\includegraphics[width=\textwidth]{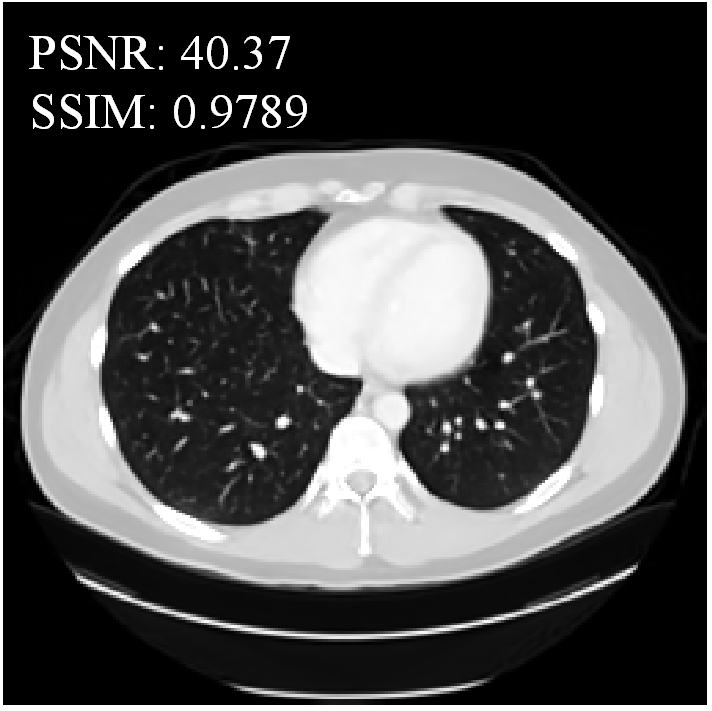}
	\centerline{(d)}
	\end{minipage} \\
         \begin{minipage}[t]{0.22\textwidth}
	\centering
	\includegraphics[width=\textwidth]{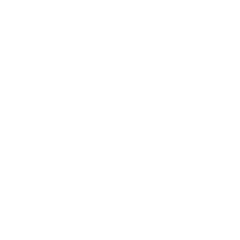}
	% \centerline{(d)}
	\end{minipage}
          \begin{minipage}[t]{0.22\textwidth}
	\centering
	\includegraphics[width=\textwidth]{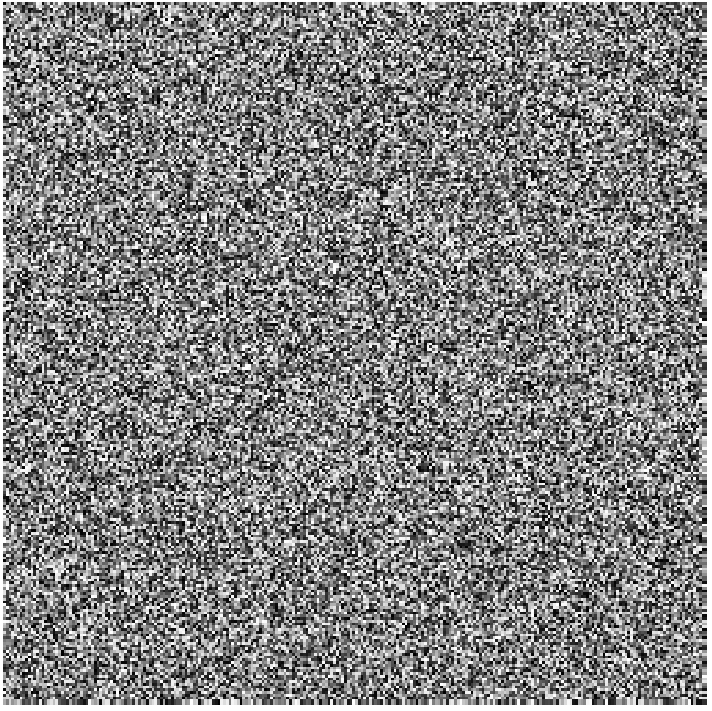}
	\centerline{(e)}
	\end{minipage} 
          \begin{minipage}[t]{0.22\textwidth}
	\centering
	\includegraphics[width=\textwidth]{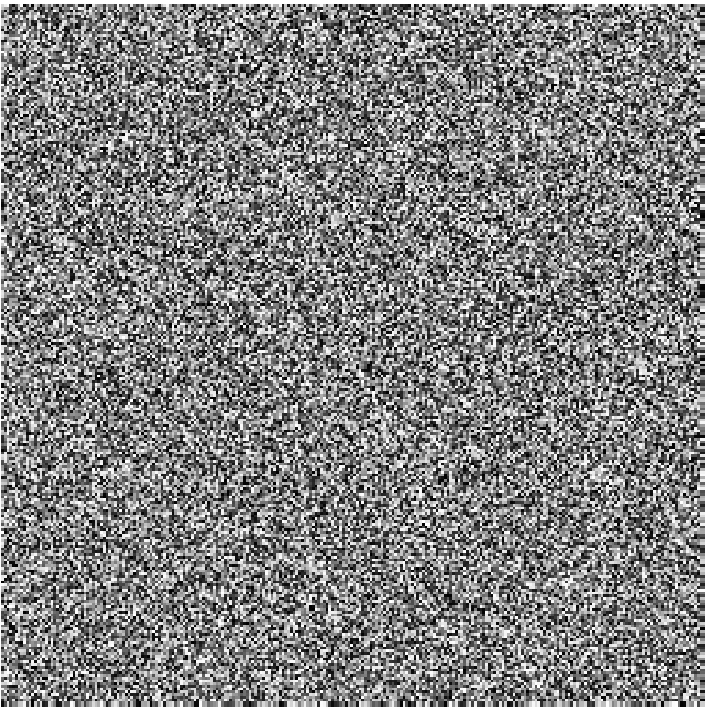}
	\centerline{(f)}
	\end{minipage} 
          \begin{minipage}[t]{0.22\textwidth}
	\centering
	\includegraphics[width=\textwidth]{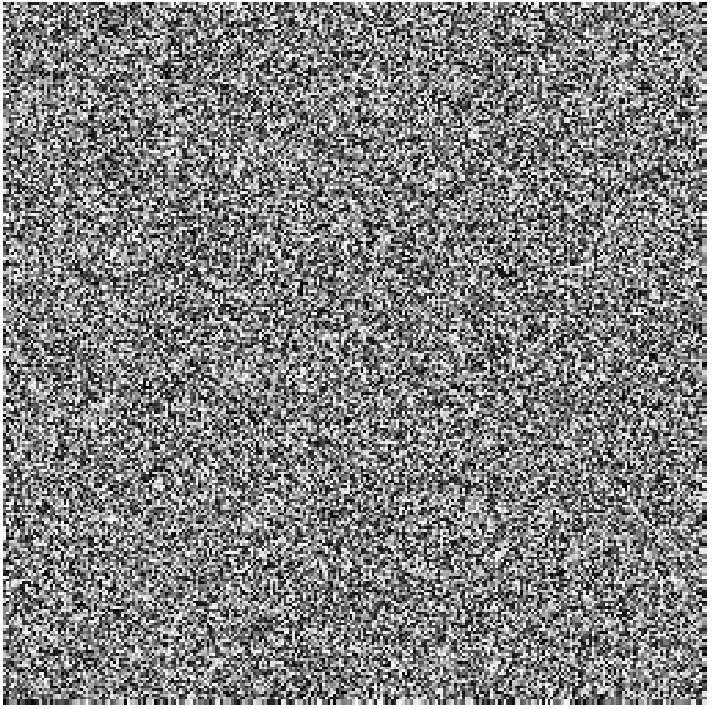}
	\centerline{(g)}
	\end{minipage} 
 \vspace{-5pt}
 \caption{The examples of plaintexts and ciphertexts. (a) represents the ground truth. (b)-(d) represent the input, denoised image with linear model and denoised image with nonlinear model, respectively. (e-g) represent the ciphertext corresponding to (b)-(d), respectively.}
 \vspace{-5pt}
 \label{fig:result}
\end{figure*}

Then we prove the correctness of the reformulated multiplication operation in the encrypted domain, as shown in Eq.~\eqref{eq:conv}. As mentioned earlier, the parameters of the model are plaintext values, and the features are encrypted. To simplify, we assume that the parameter as $a$ and $c=Enc(m)$, then we can obtain:

\begin{equation}
\begin{aligned}
    c_{mul}&=c^a \bmod n^2=g^{ma}r^{n}\bmod n^2.
\end{aligned}
\end{equation}

Then, we can decrypt $c_{mul}$ as:
\begin{equation}
    \begin{aligned}
        Dec(c_{mul})&=L(c^\lambda_{mul}\bmod n^2)\cdot \mu \bmod n \\
        &= L(g^{ma\lambda}r^{n\lambda}\bmod n^2) \cdot \mu \bmod n \\
        &= L(g^{ma\lambda}\bmod n^2) \cdot \mu \bmod n \\
        &= ma.
    \end{aligned}
\end{equation}

% Based on the We can simplify the above equation as follows:
% \begin{equation}
%     \begin{aligned}
% Dec(c_{mul})&=L(g^{ma\lambda}\bmod n^2) \cdot \mu \bmod n
% &= ma \bmod n.
%     \end{aligned}
% \end{equation}

So far, we have obtained the decryption of $c_{mul}$ corresponding to the multiplication of the plaintext $m$ by the scalar $a$. 

In this subsection, we have proved the correctness of our reformulated addition and multiplication operations in the encrypted domain, which serves as the foundation for the seamless integration of our approach within the broader context of privacy-preserving computations.

% As a result, we have successfully provid the correctness proof for the two reformulated mathematical operations.

\section{Experiments and discussions}
\label{sec:exp}
\subsection{Experimental Setting}
The “2016 NIH-AAPM-Mayo Clinic Low-Dose CT Grand Challenge” dataset~\cite{mccollough2016tu} was used to evaluate the proposed method. Poisson noise and electronic noise were added to the measured projection data to simulate the low-dose cases, which can be formulated as follows:
\begin{equation}
      y=\ln \frac{{d}_{0}}{\operatorname{Poisson}\left({d}_{0} \exp (-\hat{y})\right)+\operatorname{Normal}\left(0, \sigma_{e}^{2}\right)},
\end{equation}
where $\hat{y}$ represents the noisy-free projection, and $\sigma^2_e$ is the variance of electronic noise, which is set to 10 according to~\cite{xia2021ct}. ${d}_{0}$ denotes the photon number of incident X-rays and in this paper, and ${d}_{0}=1e^6$ is treated as the normal dose following~\cite{niu2014sparse}.

There are seven key parameters to affect the final imaging quality, including the number of detector bins, pixel length, detector bin length, the distance between the source and rotation center, the distance between the detector and rotation center as well as the photon number of incident X-rays. Related settings are set to 384, 350, 1.4 mm, 2.5 mm, 500 mm, 300 mm, and 1.25$e^5$, respectively. The well-known RED-CNN~\cite{chen2017low} was used as the denoising network, which was optimized with Adam optimizer~\cite{kingma2014adam}, with a learning rate of $1e^{-5}$. RED-CNN was implemented in PyTorch, and our reformulated operations were implemented in Numpy. Experiments were performed on an NVIDIA GTX 3080Ti GPU and AMD R5 3600 CPU.

% \begin{table}[]
% \centering
% \caption{The Parameters of Geometries and Dose Levels.}
% \label{table_setting}
% \resizebox{0.55\columnwidth}{!}{\begin{tabular}{ccc}
% % \begin{tabular}{@{}lllllllllll@{}}
% \hline
% \centering
% & PSNR & SSIM\\ \hline
% w/o FL $\dagger$ & 38.47 & 0.9448 \\
% w/o FL $\ddagger$  & 40.02 & 0.9600\\ 
% FedAvg $\dagger$  & 40.29 & 0.9609\\
% FedAvg $\ddagger$ & 40.19 & 0.9581\\ 
% FedProx $\dagger$  & 39.27 & 0.9527\\
% FedProx $\ddagger$ & 39.57 & 0.9535\\ \hline
% HyperFed & 41.20 & 0.9638\\\hline
% \end{tabular}}
% \end{table}

\subsection{Denoising Experiments}

In this subsection, we show examples of encrypted CT images, as depicted in Fig. \ref{fig:result}. The visual results demonstrate that our encryption process effectively conceals private information within the ciphertext.
Furthermore, we illustrate the decrypted examples of LDCT images based on our interactive frameworks for both linear and nonlinear models, and the qualitative and quantitative results are presented in Fig. \ref{fig:result}. These results reveal that our method effectively reduces noise using both linear and nonlinear models even if the models cannot access any plaintext features. 

Besides, it can be noticed that the performance of linear models is not as good as nonlinear models due to the lack of nonlinear operations in the network, but its communication overhead is relatively small. Specifically, the communications costs are summarized in Tab. \ref{tab:com}. In Tab. \ref{tab:com}, it can be seen that for linear models, the upload and download contents only consist of the LDCT image $x_l$ and the denoised CT image $\hat{x}$. Nonlinear models involve additional external communication overhead for intermediate features, and the total external overhead is $\sum_{i=1}^k\mathcal{C}^i$.

\begin{figure}
        \begin{minipage}[t]{0.4\columnwidth}
	\centering
	\includegraphics[width=\columnwidth]{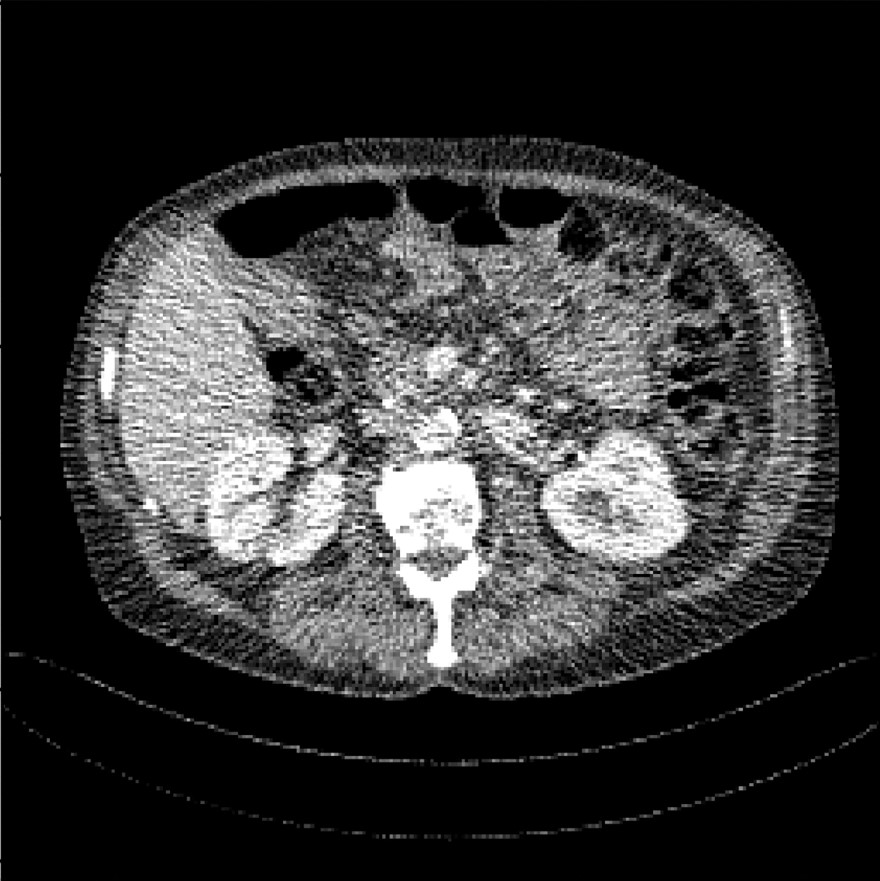}
	\centerline{(a)}
	\end{minipage}
         \begin{minipage}[t]{0.4\columnwidth}
	\centering
	\includegraphics[width=\columnwidth]{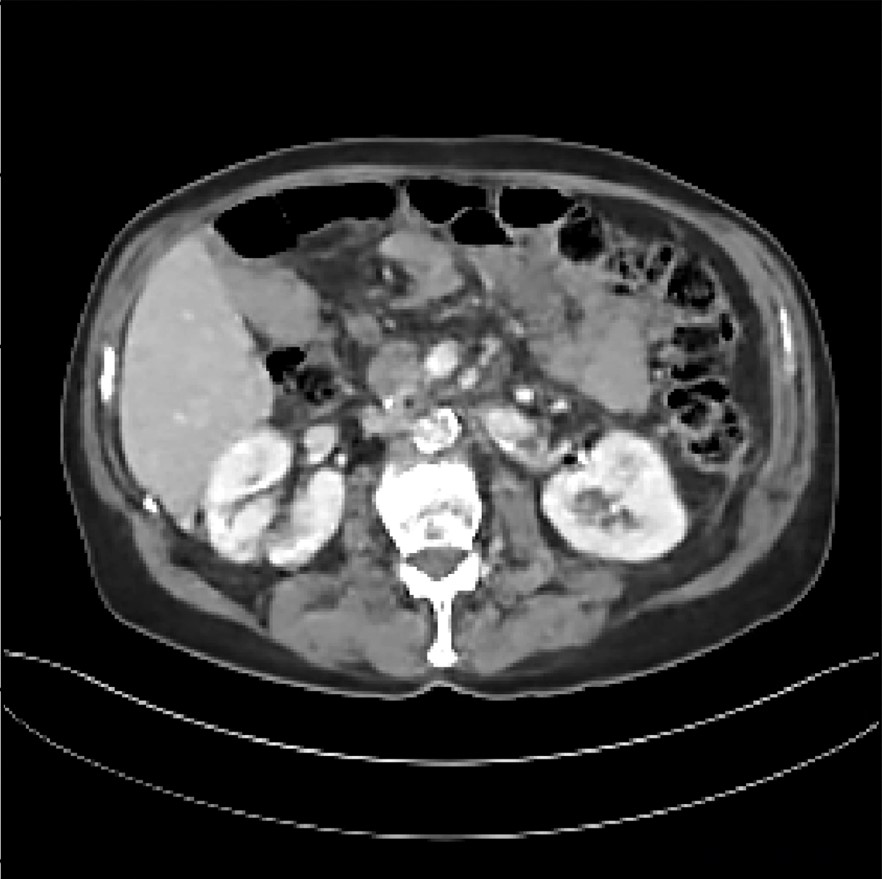}
	\centerline{(b)}
	\end{minipage}
\\
         \begin{minipage}[t]{0.4\columnwidth}
	\centering
	\includegraphics[width=\columnwidth]{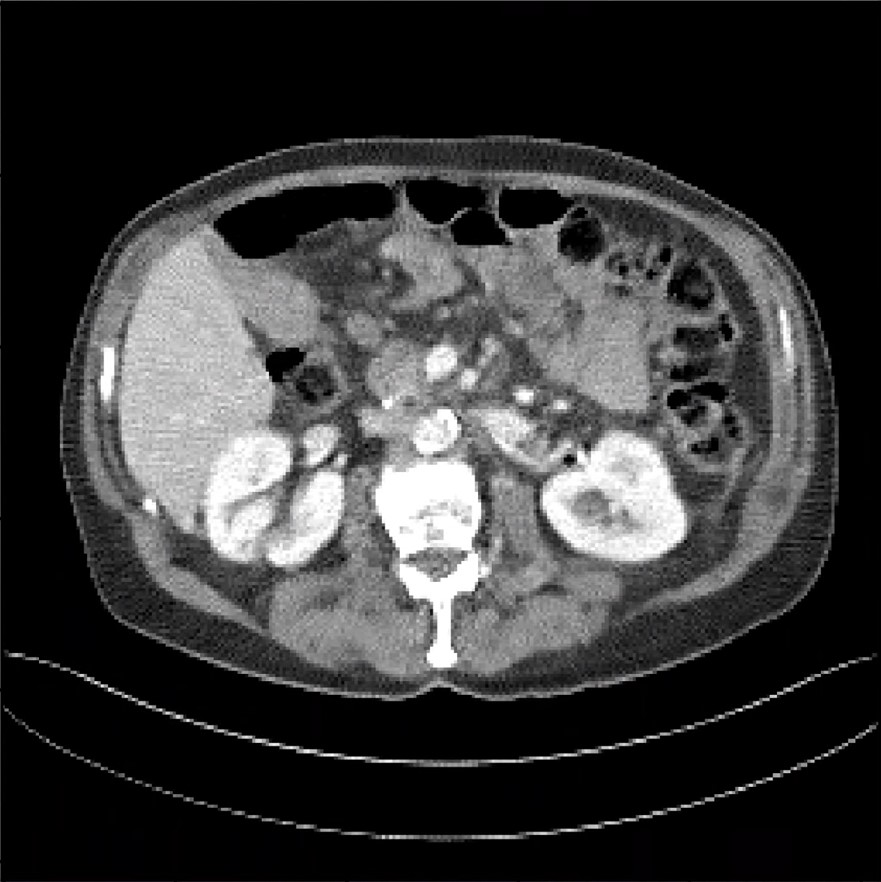}
	\centerline{(c)}
	\end{minipage}
          \begin{minipage}[t]{0.4\columnwidth}
	\centering
	\includegraphics[width=\columnwidth]{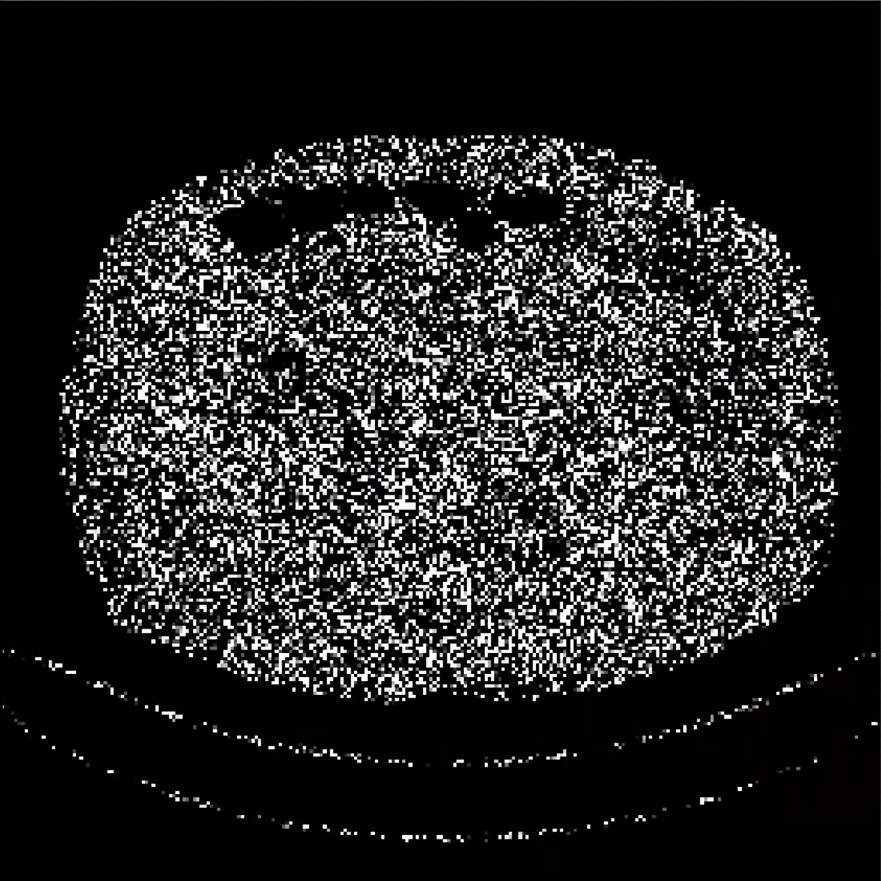}
	\centerline{(d)}
	\end{minipage} 
\vspace{-10pt}
 \caption{The results of attack experiment. (a) represents the LDCT image. (b) represents the denoised image. (c) represents the attack result from clean features. (d) represents the attack result from perturbed features. }
 \label{fig:attack}
\end{figure}

\begin{table}[]
    \centering
        \caption{The Communication Cost for Nonlinear and Linear Models.}
    \label{tab:com}
    \begin{tabular}{ccc}
    \hline
     & Upload Overhead & Download Overhead\\ \hline
        Nonlinear Model & $x_l + \sum_{i=1}^k\mathcal{C}^i$ & $\hat{x}$ \\
        Linear Model & $x_l$ & $\hat{x}$   \\ \hline
    \end{tabular}
\end{table}

\subsection{Attack Experiments}

In this subsection, we will verify the security considerations regarding the implementation of the proposed interactive framework for nonlinear models. An important step in our framework is to transfer encrypted and perturbed intermediate features, as shown in Eq.~\eqref{eq:disturb}. Hence, there arises a privacy concern about this operation. One intuitional question is: "Is it possible for a malicious user to decrypt the features and design a model to steal the knowledge of the server model?" In theory, if the user can steal knowledge of a single layer of the network, they could repeat this attack iteratively, eventually compromising the entire intellectual property of the service model.

To address related concerns, we conduct attack experiments to validate the security of the proposed framework. In these experiments, we assume that hackers can access the decrypted intermediate features. To make a challenging attack scenario, we further assume that the hackers' goal is to steal the knowledge of the last layer of the server model for a single LDCT image. We illustrate the results of these attacks in Fig.~\ref{fig:attack}. 
The results indicate that the hacker can indeed learn knowledge about the server model from the clean features. However, our perturbation operation effectively thwarts such attacks, even when the target is a single image and the perturbance matrix is fixed. In practice, our perturbance matrix is randomly sampled, significantly increasing the difficulty of the attack. Hence, our method can effectively defend against this attack.

% The image shows that after the hacker obtains the original features and results, he can train an attack network to attack the server network parameters, resulting in parameter leakage. However, using the disturbed features, the network model is well protected, and hackers cannot obtain a network that generates similar results.

% \vspace{-20pt}
\section{Conclusion}
\label{sec:con}
In this paper, we propose a novel interactive framework for LDCT denoising, enabling the model to work directly in the encrypted domain without any extra operations. Our method offers two significant advantages: preserving patients' privacy and safeguarding the intellectual property of the server model. We give mathematical proof and attack experiments to validate the correctness and security of our method. This work presents a promising solution for privacy concerns amid the surge in large models, especially in medical tasks. However, challenges still exist, such as software compatibility. Leading DL frameworks like PyTorch and TensorFlow lack support for large numbers which are crucial for cryptographic security. Thus, future research must extend DL frameworks to accommodate these large numbers.

\bibliographystyle{ACM-Reference-Format}
\bibliography{ref}

% \appendix









\end{document}